\documentclass[printer]{tMOP2eJV}
\usepackage{epstopdf} % needed for figures

\newcommand{\beq}{\begin{equation}}
\newcommand{\eeq}{\end{equation}}
\newcommand{\beqa}{\begin{eqnarray}}
\newcommand{\eeqan}{\end{eqnarray*}}
\newcommand{\beqan}{\begin{eqnarray*}}
\newcommand{\eeqa}{\end{eqnarray}}
\newcommand{\bra}[1]{\langle{#1}|}
\newcommand{\ket}[1]{|{#1}\rangle}
\newcommand{\ip}[1]{\langle{#1}\rangle}

\newcommand{\Tr}{{{\rm Tr}}}

\begin{document}

%\doi{10.1080/09500340.2013.812252}
% \issn{1362-3044}
%\issnp{0950-0340} \jvol{00} \jnum{00} \jyear{2012} \jmonth{5 September}

\markboth{J.A. Vaccaro}{Comment on the Wigner phase distribution}

%\articletype{PREPRINT}

\title{\vspace{-4.5cm}Comment on `Operator formalism for the Wigner phase distribution'}

\author{Joan A. Vaccaro$^{\dagger}$\thanks{$^\dagger$Email: J.A.Vaccaro@griffith.edu.au}\\
\vspace{6pt} Centre for Quantum Computation and Communication Technology
(Australian Research Council), Centre for Quantum Dynamics, Griffith
University, 170 Kessels Road, Nathan 4111, Australia
}

\maketitle

\begin{abstract}
The operator associated with the radially integrated Wigner function is found
to lack justification as a phase operator.

\bigskip


\end{abstract}

\section{Introduction}

The radially integrated Wigner function has been shown by Garraway and Knight
\cite{Garraway1, Garraway2} to be negative for important classes of states.
As such, it cannot be taken to represent physical properties in the way that
probability distributions represent the statistical properties of the things
they describe. So it comes as something of a surprise to find the same
function being used as a basis for describing the phase observable of a
single mode radiation field, as Subeesh and Sudhir have recently done
\cite{Subeesh}. They derive an operator $\hat\rho_W(\theta)$,
\beq \label{rho_W}
   \hat\rho_W(\theta) = \frac{1}{2\pi}\sum_{m,n=0}^\infty\sum_{l=0}^n
   \frac{(-1)^m 2^{m+n/2}\exp[i(n-2l)\theta]}{m!(n-l)!l!}
    \Gamma\left(\frac{n}{2}+1\right){\hat a}^{\dagger m+n-l}{\hat a}^{m+l} \ ,
\eeq
which they call the Wigner phase operator, whose expectation value is the radially integrated
Wigner function $P^W_\psi(\theta)$, i.e.
\beq
   P^W_\psi(\theta)=\Tr[\hat\rho_\psi\hat\rho_W(\theta)]
\eeq
for system density operator $\hat\rho_\psi$. Calling $\hat\rho_W(\theta)$ a
{\em phase operator} is to lay claim that it plays a meaningful role in
describing phase. Indeed, the final paragraph of \cite{Subeesh} states that
``the radially integrated Wigner function captures essentially the same phase
information'' as the Pegg-Barnett phase formalism. Such extraordinary claims,
if left unqualified, may lead the unwary to the conclusion that the operator
$\hat\rho_W(\theta)$ does indeed faithfully represent quantum phase. The
purpose of this comment is to point out that it does not.

\begin{figure}  %%%%%%%%%%%%%%%%%%%%%%%%%%%%%%%%%%%%%
\begin{center}
  \includegraphics[width=7cm]{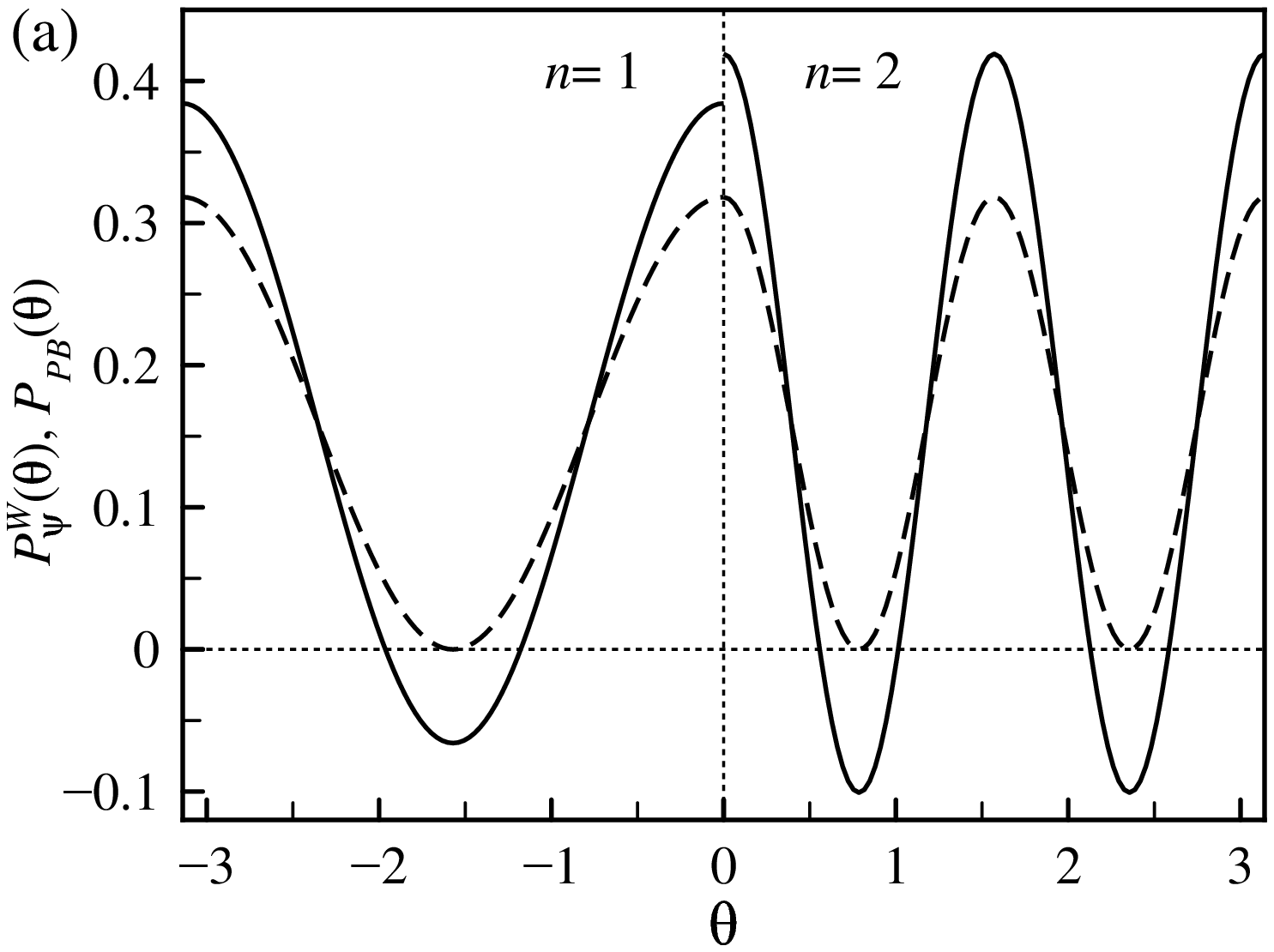} \includegraphics[width=7cm]{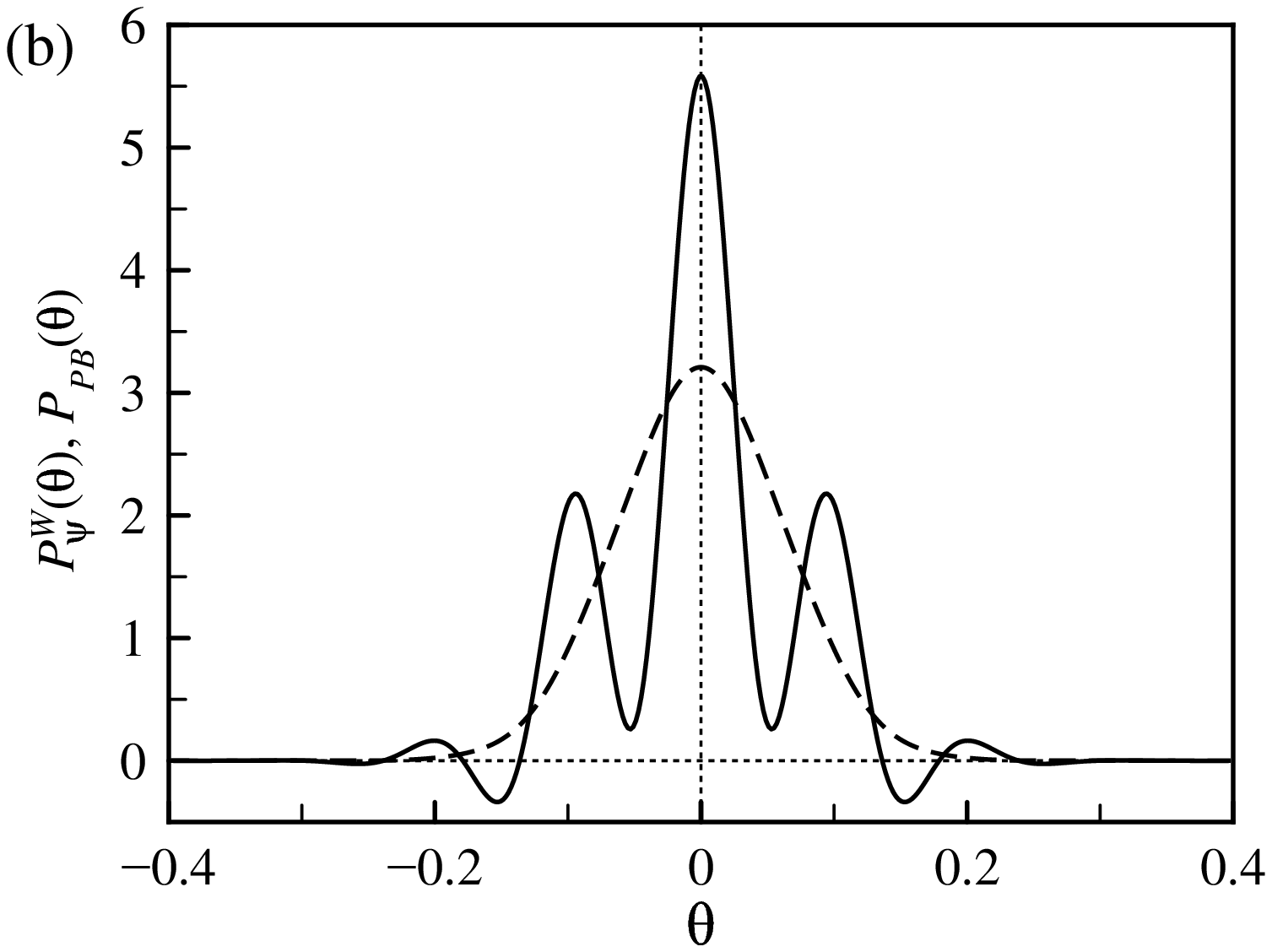}

  \caption{Comparisons of the
  functions $P^W_\psi(\theta)$ (solid curve) and $P_{PB}(\theta)$ (dashed
  curve) for various states.  In (a) the state is  $(\ket{0}+\ket{2n})/\sqrt{2}$ where, for clarity, only the $\theta<0$ segment is plotted for $n=1$ and the $\theta>0$ segment for $n=2$.  In (b) the state is $c(\ket{\alpha}+\ket{\beta})$ where $c$ is a normalisation constant and $\alpha=-2$ and $\beta=8$.
\label{fig1}}
\end{center}
\end{figure}    %%%%%%%%%%%%%%%%%%%%%%%%%%%%%%%%%%%%%

It is not difficult to find grounds to counter the claims of \cite{Subeesh}. Foremost is the
work of Garraway and Knight \cite{Garraway1,Garraway2} that shows
$P^W_\psi(\theta)$ differs significantly from the Pegg-Barnet phase
probability density $P_{PB}(\theta)$ \cite{PB1,PB2,PB3}. In particular, Garraway and Knight
have shown that $P^W_\psi(\theta)$ can have negative values for wide classes
of states whereas $P_{PB}(\theta)$, being a genuine probability distribution,
is always nonnegative. The amount of negativity in $P^W_\psi(\theta)$ can be
significant. For example, Figure 1 (a) illustrates the negativity of
$P^W_\psi(\theta)$ for the state $(\ket{0}+\ket{2n})/\sqrt{2}$, for which
\[
 P^W_\psi(\theta) = \frac{1}{2\pi}\left[1+\frac{2^n n!}{\sqrt{(2n)!}}\cos(2n\theta)\right]
\]
whereas, for comparison,
\[
   P_{PB}(\theta)=\frac{1}{2\pi}[1+\cos(2n\theta)]
\]
 is everywhere nonnegative. While $P^W_\psi(\theta)$ and $P_{PB}(\theta)$ do share some similarities for special classes of states, notably number and coherent states
\cite{Garraway1,Garraway2,Subeesh}, their significant differences for other
states is sufficient to quash any notion that they capture the same
information about phase in any general sense. An example of how different they can be is given by Figure 1 (b) which compares $P^W_\psi(\theta)$ and $P_{PB}(\theta)$ for the superposition of coherent states $c(\ket{\alpha}+\ket{\beta})$, where $c$ is a normalisation constant and $\alpha=-2$ and $\beta=8$.
Moreover, while
$P_{PB}(\theta)$ is a probability density that describes the outcome of an
ideal measurement of phase, the same cannot be said for $P^W_\psi(\theta)$
whose negativity {\it forbids} any interpretation as a probability associated with
a measurement, on principle. The claim that $P^W_\psi(\theta)$ represents
phase properties {\it per se} is therefore unjustified.

There are also difficulties associated with the operator $\hat\rho_W(\theta)$ defined in \cite{Subeesh}.  It is immediately apparent from the
negativity of $P^W_\psi(\theta)$ that $\hat\rho_W(\theta)$ is {\em not} a
positive operator-valued measure. It follows that it is not a projection
operator and so it does not project onto states of well defined phase.
However, it does have the property that it undergoes phase shifts in the
sense that \cite{archive}
\[
   \hat\rho_W(\theta)=e^{i\hat N\theta}\hat\rho_W(0) e^{-i\hat N\theta}
\]
where $\hat N=\hat a^\dagger \hat a$ is the number operator.  This property
underlies the relationship in equation (14) of \cite{Subeesh},
\[
      \Tr[\hat \rho_W(\theta)\hat\rho_{PB}(\phi)]=\Tr[\hat \rho_W(\theta')\hat \rho_{PB}(\phi')]
\]
for $\theta-\phi=\theta'-\phi'$, that the authors call ``weak-equivalence''
between $\hat\rho_W(\theta)$ and the Pegg-Barnett phase operator
$\hat\rho_{PB}(\theta)=\ket{\theta}\bra{\theta}$ \cite{weak limit}. But there
are many operators that are weakly equivalent to $\hat\rho_{PB}(\theta)$ in
the same sense. For example, let
\[
   \hat A(\theta)=e^{i\hat N\theta}\hat A_0 e^{-i\hat N\theta}
\]
for an arbitrary operator $\hat A_0$.  Then
\[
    \Tr[\hat A(\theta)\hat \rho_{PB}(\phi)]=\Tr[\hat A\hat(\theta') \rho_{PB}(\phi')]
\]
for $\theta-\phi=\theta'-\phi'$ and so $\hat A(\theta)$ is also weakly
equivalent to $\hat\rho_{PB}(\theta)$ in the same way. Evidently this weak
equivalence is not a stringent condition.  It certainly isn't strong enough
to justify the definition of $\hat \rho_W(\theta)$.

To illustrate the kind of caution that must be used when dealing with
$\hat\rho_W(\theta)$, consider the angle operator defined by
\beqan
    \hat Q&=&\int_{\theta_0}^{\theta_0+2\pi}\theta \hat\rho_W(\theta) d\theta
\eeqan
for arbitrary phase angle $\theta_0$.  Its expectation value for state $\hat\rho_\psi$ is given by
\[
   \ip{\hat Q}=\int_{\theta_0}^{\theta_0+2\pi}\theta P^W_\psi(\theta) d\theta
\]
in which $P^W_\psi(\theta)$ appears to play the role of a probability but,
owing to $\hat \rho_W(\theta)$ not being a projection operator, it is
straightforward to show that
\[
     \hat Q^n\ne\int_{\theta_0}^{\theta_0+2\pi}\theta^n \hat\rho_W(\theta) d\theta
\]
for integer $n>1$ and so it follows that
\[
     \ip{\hat Q^n}\ne\int_{\theta_0}^{\theta_0+2\pi}\theta^n P^W_\psi(\theta) d\theta
\]
in general.  This means that the statistical properties of the operator $\hat
Q$ are not described by $P^W_\psi(\theta)$. In contrast, it is well-known that the Pegg-Barnett hermitian phase operator $\hat\phi_{s}$,
\[
   \hat\phi_{s}=\sum_{m=0}^s \theta_m \ket{\theta_m}\bra{\theta_m}
\]
where $\theta_m=\theta_0+2\pi/(s+1)$ and $\ket{\theta_m}$ is a phase eigenstate, has the properties that
\beqan
   \hat\phi_{s}^n&=&\sum_{m=0}^s \theta_m^n \ket{\theta_m}\bra{\theta_m}\\
   \ip{\hat\phi^n}&=&\lim_{s\to\infty} \ip{\hat\phi_{s}^n}= \int_{\theta_0}^{\theta_0+2\pi}\theta^n P_{PB}(\theta) d\theta
\eeqan
where $\ip{\hat\phi^n}$ are the moments of the probability density $P_{PB}(\theta)$ \cite{PB1,PB2,PB3}.

In conclusion, the radially integrated Wigner function cannot be interpreted
as a probability owing to its negativity, and as such it cannot be used to
describe the statistics of an observable such as phase. Using it as a basis
to define the operator $\hat\rho_W(\theta)$ to represent phase is therefore
without appropriate justification.  This leaves the merits of the operator
$\hat\rho_W(\theta)$ introduced in \cite{Subeesh} open to question.

\section*{Acknowledgement}
This research was conducted by the Australian Research Council Centre of
Excellence for Quantum Computation and Communication Technology (Project no.
CE110001027).

\label{lastpage}


\begin{thebibliography}{23}
\bibitem{weak limit}
Strickly speaking, $\ket{\theta}\bra{\theta}$ is a weak limit of the
Pegg-Barnett phase state projection operator as described in \cite{PhysScr}.
\end{thebibliography}
\end{document}